\documentclass[twocolumn,letter]{jpsj3}

\usepackage{color}
\usepackage{bm}

\title{%
Delocalization of Surface Dirac Electrons
in Disordered Weak Topological Insulators
}

\author{%
Yositake Takane
}

\inst{%
Department of Quantum Matter, Graduate School of Advanced Sciences of Matter,\\
Hiroshima University, Higashihiroshima, Hiroshima 739-8530, Japan
}

\recdate{ \hspace{50mm} }

\abst{%
The spectrum of massless Dirac electrons on the side surface of
a three-dimensional weak topological insulator is significantly affected by
whether the number of unit atomic layers constituting the sample
is even or odd; it has a finite-size energy gap in the even case
while it is gapless in the odd case.
The conductivity of such a two-dimensional Dirac electron system
with quenched disorder is calculated
when  the Fermi level is located at the Dirac point.
It is shown that the conductivity increases with increasing disorder
and shows no clear even-odd difference
when the aspect ratio of the system is appropriately fixed.
From the system-size dependence of the average conductivity,
the scaling function $\beta$ is determined
under the one-parameter scaling hypothesis.
The result implies that $\beta = 0$ in the clean limit
at which the conductivity is minimized,
and that $\beta > 0$ otherwise.
Hence, the system is a perfect metal in the thermodynamic limit
except in the clean limit that should be regarded as an unstable fixed point.
}

\begin{document}
\maketitle

Three-dimensional topological insulators are characterized by
four $\mathbb{Z}_2$ indices, the strong index $\nu_0$
and the weak indices $\nu_1$, $\nu_2$, and $\nu_3$.~\cite{fu,moore,roy}
The strong index distinguishes a strong topological insulator (STI)
with $\nu_0 = 1$ from a weak topological insulator (WTI)
and a trivial insulator with $\nu_0 = 0$.
Among the sectors of $\nu_0 = 0$, a WTI is characterized by the condition
that at least one of the weak indices is nonzero.
The most notable feature of STI and WTI is that low-energy electron states
described by a massless Dirac equation appear on their surfaces.
They are called Dirac electrons and possess
a linear energy dispersion forming a gapless conic structure
(i.e., Dirac cone) in the reciprocal space.
A crucial difference between STI and WTI is that
Dirac electrons appear on every surface of a sample in the strong case
while, in the weak case, they disappear on the surface normal to
the weak vector $\mib{\nu} \equiv (\nu_1,\nu_2,\nu_3)$.
Noting that WTIs have a layered structure and $\mib{\nu}$ designates
the direction along which unit atomic layers are stacked,
we see that Dirac electrons in a WTI are confined on its side surface.
Another important difference is that an STI typically has one Dirac cone,
while a WTI has two Dirac cones.
Thus, the surface state of a WTI was considered to be weak against disorder,
getting gapped by scattering between two Dirac cones.
However, it has been shown that a WTI is not necessarily weak.~\cite{ran,
imura1,ringel,mong,liu1,imura2,yoshimura,kobayashi,morimoto,obuse1}
Several materials have been proposed
as candidates for a WTI.~\cite{yan,rasche,tang,yang}

Dirac electrons on the side surface of a WTI show unusual properties
that depend on whether the number of unit atomic layers
stacked along $\mib{\nu}$ is even or odd.~\cite{ringel,imura2,yoshimura}
Note that the unit atomic layer can be regarded as a two-dimensional (2D)
quantum spin Hall insulator
possessing a one-dimensional helical edge channel.~\cite{fu}
A series of helical edge channels forms
a Dirac electron system on the side surface.
If the number of unit atomic layers is even, the helical edge channels
acquire a finite-size gap as a result of their mutual coupling.
Contrastingly, in the odd case, one helical channel survives
without being gapped out.
Another important property arises from the fact that regardless of
the location of the Fermi level, the number of conducting channels
is even in the even case and odd in the odd case.
This parity leads to a notable difference in the transport property
of disordered systems.
From a symmetry viewpoint, Dirac electrons have symplectic symmetry
that preserves the time-reversal symmetry
without the spin-rotation invariance.~\cite{suzuura}
It has been shown that a disordered system with symplectic symmetry has
a perfectly conducting channel in the odd-channel case,~\cite{ando1}
and that this crucially affects the transport property
in a quantum wire structure;~\cite{takane1,takane2,ando2,sakai,comment1}
the system with even channels becomes insulating in the long-wire limit
while that with odd channels remains metallic.
This observation holds for the Dirac system
on the side surface of a WTI.~\cite{ringel}

Let us consider a 2D Dirac electron system on the surface of WTIs,
focusing on its transport properties in the large area limit.
An important question is
whether it is localized by quenched disorder.~\cite{ringel}
In the case of an STI, it has been demonstrated that the 2D Dirac system
with one Dirac cone is a perfect metal.~\cite{bardarson,nomura}
The case of a WTI with two Dirac cones has been studied by
Mong {\it et al.}~\cite{mong} and Obuse {\it et al.}~\cite{obuse1}
using a finite-size scaling approach.~\cite{abrahams}
Both of them conclude that the system is again a perfect metal
showing no sign of Anderson localization.
However, there remain subtle issues to resolve.
Mong {\it et al.} derived a scaling relation of the conductivity
on the basis of a microscopic model consisting of two Dirac cones.
This model is plausible but involves no even-odd feature
since it is defined on a 2D continuous space.
Obuse {\it et al.} elaborately studied the localization problem
by using a network model~\cite{obuse1,obuse2} in which
the presence of a perfectly conducting channel is encoded.
However, they did not explicitly present a scaling relation of
the conductivity including its parity dependence.

In this letter, we numerically study the localization problem on the surface
of a WTI using a microscopic model that correctly describes
the even-odd difference.
Our attention is focused on the case of the Fermi level
being located at the Dirac point~\cite{bardarson}
since the conductivity is expected to be minimized there.~\cite{shon,titov}
We numerically calculate the average conductivity of disordered systems
of length $L$ and width $W$ at a fixed ratio $R \equiv L/W$,
and analyze its behavior using a finite-size scaling approach.
We separately treat the even and odd cases.
It is shown that the $L$ dependence of the average conductivity becomes
parity-independent if $R$ is sufficiently small.
It is also shown that the average conductivity increases with increasing
disorder and is minimized in the clean limit.
From the numerical result, we determine the scaling function $\beta$
under the one-parameter scaling hypothesis.~\cite{abrahams}
The result implies that $\beta > 0$ except in the clean limit
that should be regarded as an unstable fixed point of $\beta = 0$.
That is, the system becomes a perfect metal with increasing system size,
except at the unstable fixed point.
We set $\hbar = 1$ throughout this letter.

We consider the side surface of width $W$ on the $xy$-plane
being infinitely long in the $x$-direction.
It consists of $M$ unit atomic layers stacked in the $y$-direction
and the width is $W = Ma$ with $a$ being the lattice constant.
Each unit layer has one helical edge channel aligned with the $x$-axis,
and the resulting $M$ helical channels form the 2D Dirac system
by coupling with their nearest neighbors.
The region of $L \ge x \ge 0$ is regarded as the sample with disorder,
and the region of $x < 0$ ($x > L$) plays a role of the left (right) lead.
Let $|j\rangle \equiv \{|j\rangle_{\uparrow},|j\rangle_{\downarrow}\}$
be the two-component vector representing the state
in the $j$th helical channel ($M \ge j \ge 1$),
where $\uparrow$ and $\downarrow$ specify the spin direction.
It is assumed below that the up-spin (down-spin) state propagates
in the right (left) direction.
We adopt the following effective Hamiltonian
for the surface state of a WTI:~\cite{morimoto,obuse1}
\begin{align}
       \label{H_eff}
   H_{\rm eff}
 & = \sum_{j=1}^{M}
     |j\rangle \left[ \begin{array}{cc}
                          -iv\partial_{x}+U(x) & 0 \\
                          0 & iv\partial_{x}+U(x)
                        \end{array} \right]
     \langle j|
   \nonumber \\
 &  \hspace{0mm}
   + \sum_{j=1}^{M-1}
     \left\{
     |j+1\rangle \left[ \begin{array}{cc}
                          0 & \frac{v}{2a} \\
                          -\frac{v}{2a} & 0
                          \end{array} \right]
     \langle j|
   + {\rm h.c.}
     \right\} .
\end{align}
The potential $U(x)$ is included to simulate the setup~\cite{katsnelson}
in which the Fermi level is fixed at the Dirac point ($\epsilon = 0$)
in the sample region, while the right and left leads are deeply doped.
That is, we set $U(x) = 0$ in the sample region,
and $U(x) = - U_0$ outside the sample with $U_0$ being positive and large.

Let us briefly describe the wave functions at an energy
$\epsilon$ in the case of $U(x) \equiv 0$.
Our model has two Dirac cones centered at $(0,0)$ and $(0,\pi/a)$
in the reciprocal space.
The transverse function is constructed by superposing
two wave functions of different Dirac cones as~\cite{imura2}
\begin{align}
    \label{eq:transv-f}
  \chi_{m}(j) = c_{m} \left(e^{iq_{m}y_j}-(-1)^{j}e^{-iq_{m}y_j} \right)
\end{align}
with $y_j \equiv ja$, where $c_m$ is the normalization constant and
$q_{m} = m\pi/[(M+1)a]$ with
\begin{align}
  m = \frac{M-1}{2}, \frac{M-3}{2}, \dots,-\frac{M-1}{2} .
\end{align}
The subband energy of the $m$th mode is $\Delta_{m} = (v/a)|\sin q_{m}a|$,
and the dispersion relation as a function of the longitudinal wave number $k$
is $\epsilon_{m}(k) = \pm \sqrt{(vk)^{2}+\Delta_{m}^2}$.
For an odd $M$, we see that $\Delta_{m}$ vanishes for $m = 0$,
indicating that the system has gapless excitations.
For an even $M$, $m = 0$ is not allowed
so a finite-size gap opens across the Dirac point (see Fig.~1).
\begin{figure}[btp]
\begin{center}
\includegraphics[height=3.9cm]{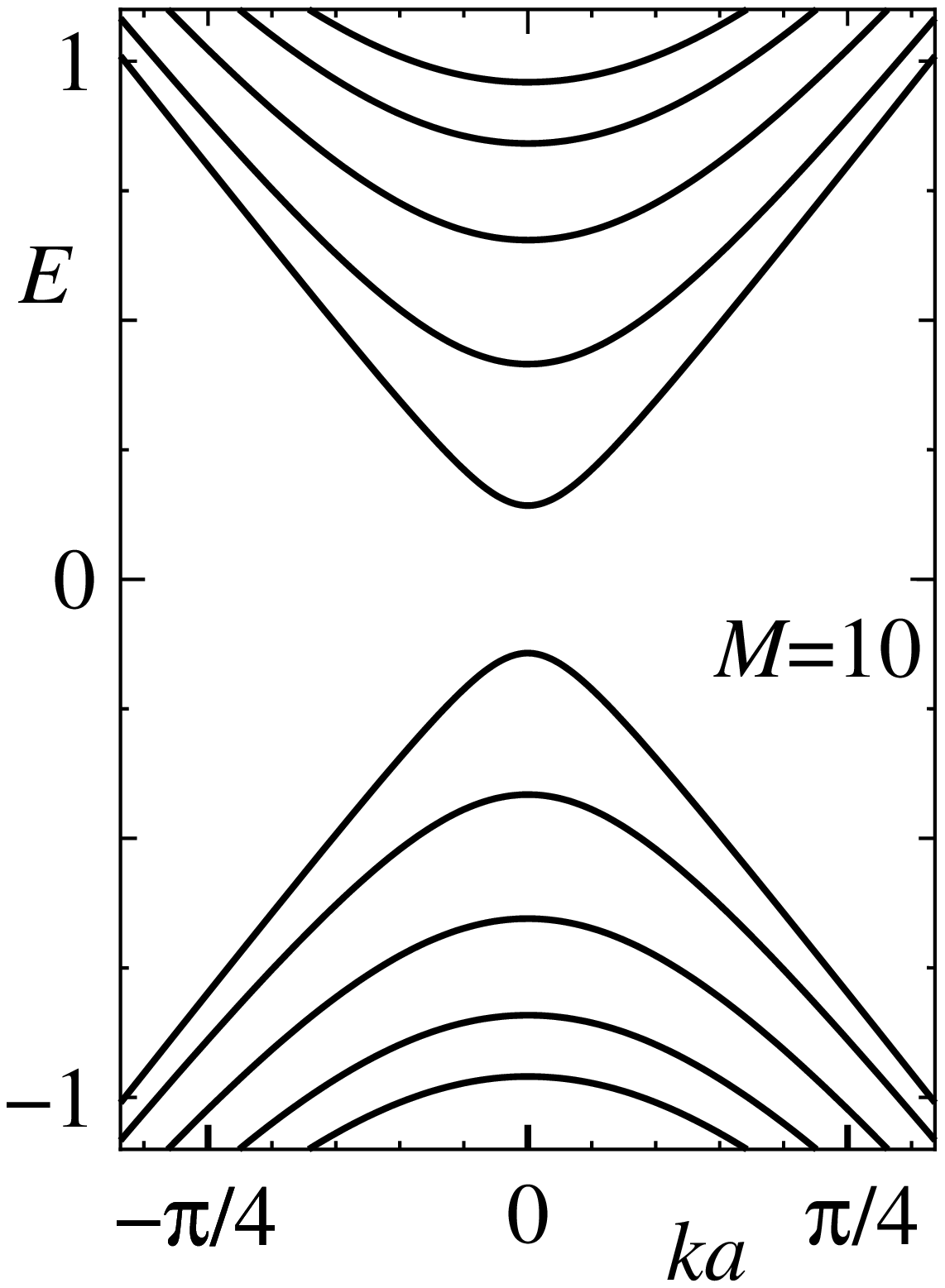}
\includegraphics[height=3.9cm]{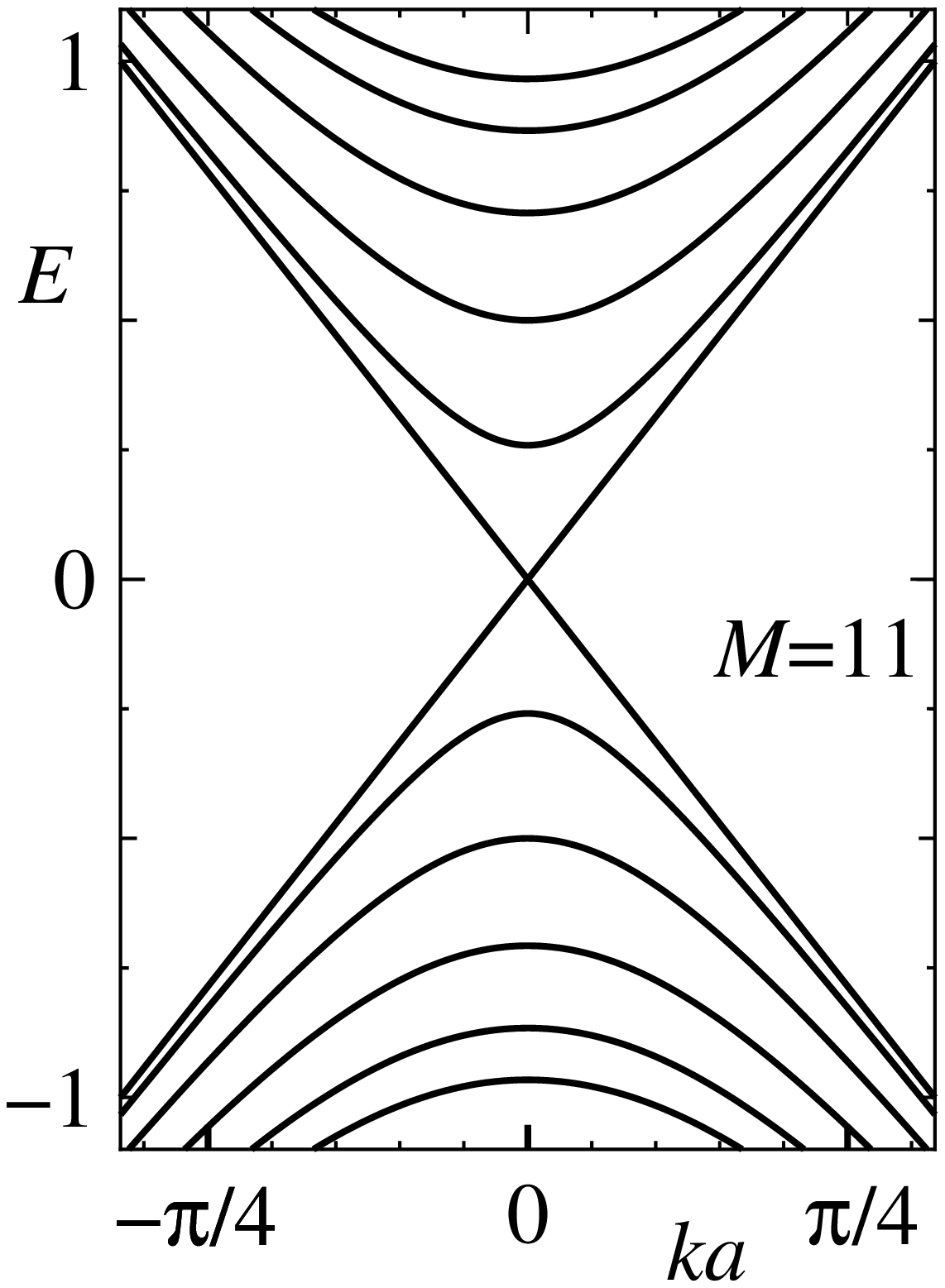}
\end{center}
\caption{Dispersion relations for $M = 10$ and $11$,
where $E$ is the normalized energy defined by $E = (a/v)\epsilon$.
}
\end{figure}
If $|\epsilon| > \Delta_{m}$,
the $m$th mode provides two counterpropagating channels.
The corresponding wave functions $\varphi_{m}^{\pm}(x,j)$ are
\begin{align}
      \label{eq:wf-propa}
   \varphi_{m}^{\pm}(x,j)
   = \frac{1}{\sqrt{v_{m}}} \chi_{m}(j)e^{\pm ik_{m}x}
     \left[ \begin{array}{c}
              a_{m}^{\pm} \\
              b_{m}^{\pm}
            \end{array} \right] ,
\end{align}
where $\pm$ specifies the propagating direction,
$k_{m} = \sqrt{\epsilon^2-\Delta_{m}^2}/v$, and
\begin{align}
   \left[ \begin{array}{c}
            a_{m}^{\pm} \\
            b_{m}^{\pm}
          \end{array} \right]
   = \frac{1}{\sqrt{|\pm vk_{m}-\epsilon|^2+\Delta_{m}^2}}
     \left[ \begin{array}{c}
            i\eta_{m}\Delta_{m} \\
            \pm vk_{m}-\epsilon
          \end{array} \right] ,
\end{align}
where $\eta_{m} = 1$ for $m > 0$ and $-1$ for $m < 0$.
The group velocity $v_{m}$ is obtained as $v_{m} = v(vk_{m}/|\epsilon|)$.
In the case of $|\epsilon| < \Delta_{m}$,
the $m$th mode provides two evanescent channels.
The corresponding wave functions are obtained
from Eq.~(\ref{eq:wf-propa}) by the following replacement:
$k_{m} \to i\kappa_{m}$ with $\kappa_{m} = \sqrt{\Delta_{m}^2-\epsilon^2}/v$.
The group velocity has no physical meaning for evanescent channels.

To incorporate disorder, we add the potential $V$
consisting of $\delta$-function-type impurities in the sample region,
given by $V = \sum_{j=1}^{M} V_{\rm imp}(x,j)|j\rangle \langle j|$ with
\begin{align}
  V_{\rm imp}(x,j)
  = \sum_{p=1}^{N_{\rm imp}}V_{p}a\delta(x-x_{p})\delta_{j,j_{p}} ,
\end{align}
where $V_p$ is the strength of the $p$th impurity located at $x = x_p$
on the $j_p$th channel, and $N_{\rm imp}$ is the total number of impurities.
The strength of disorder is characterized by the parameter $\Gamma$
defined by~\cite{bardarson}
\begin{align}
  \Gamma = \frac{a}{v^2}\sum_{j'=1}^{M}
           \int dx' \langle V_{\rm imp}(x,j)V_{\rm imp}(x',j') \rangle ,
\end{align}
where $\langle \cdots \rangle$ represents the disorder average.
If $V_p$ is assumed to be uniformly distributed within the interval of
$[-V_{0},+V_{0}]$, the parameter is evaluated as
\begin{align}
  \Gamma  = \frac{V_{0}^{2}}{(v/a)^{2}}\frac{N_{\rm imp}}{MN} ,
\end{align}
where $N$ is the dimensionless system length defined by $N \equiv L/a$.
To obtain the conductivity of the sample of area $W \times L$
with impurities, we calculate the transmission matrix $\mib{t}$
through it in terms of which the dimensionless conductance $g$
is determined as $g = {\rm tr}\{\mib{t}^{\dagger}\mib{t}\}$.
The dimensionless conductivity $\sigma$ is given by $\sigma = (L/W)g$.
The transmission matrix for a given impurity configuration is
numerically determined employing the method presented in Ref.~\citen{tamura}.
In this method, the $S$ matrix for the whole system is decomposed into
single-impurity parts and free-propagating parts.
Once they are evaluated, we can construct the $S$ matrix
using a composition law.
Here, we comment on the procedure to save computational time.
If the above method is straightforwardly applied to our system,
the number of single-impurity parts is $N_{\rm imp}$ and
that of free-propagating parts is $N_{\rm imp}+1$.
To save the computational time, it is efficient to reduce these numbers.
To do so, we randomly choose $n$ points on the $x$-axis
within the sample region as $0<x_1<x_2\dots<x_n<L$
and then put an impurity on every helical channel
at each $x_i$ ($i = 1,2,\dots,n$).
The total number of impurities is $N_{\rm imp}=M \times n$,
and the number of single-impurity parts is reduced to $n$.

\begin{figure}[tbp]
\begin{center}
\includegraphics[height=2.7cm]{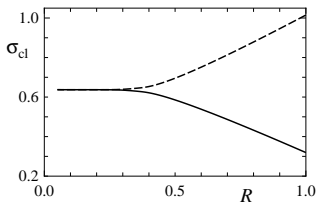}
\end{center}
\caption{Conductivity in the clean limit at $L/a = 1000$
as a function of aspect ratio $R = L/W$,
where the solid and dashed lines respectively
represent the even and odd cases.
}
\end{figure}
Before presenting the result of numerical simulations,
let us briefly consider the conductivity $\sigma_{\rm cl}$ in the clean limit,
at which the corresponding conductance $g_{\rm cl}$
is analytically obtained as~\cite{katsnelson}
\begin{align}
  g_{\rm cl}
  = \sum_{m} \cosh^{-2}\left( \Delta_{m}L/v \right) .
\end{align}
Note that the term with $m=0$ for an odd $M$
corresponds to a perfectly conducting channel.
In Fig.~2, $\sigma_{\rm cl} \equiv (L/W)g_{\rm cl}$ at $L/a = 1000$ is plotted
for the even and odd cases as a function of the aspect ratio $R = L/W$.
We see from Fig.~2 that $\sigma_{\rm cl}$ shows no even-odd difference
if $R$ is sufficiently small.
This suggests that the case of $R \ll 1$ is suitable for capturing
a parity-independent scaling flow,
while computations for a given $L$ become heavier with decreasing $R$.
Considering these two points, we set $R = 1/3$ hereafter.
Note that $\sigma_{\rm cl}(L)$ is a monotonically decreasing function of $L$
and converges in the large-$L$ limit.
At $R = 1/3$,
we find that $\sigma_{\rm cl}(\infty) \approx 0.6386$ in the odd-$M$ case
and $\sigma_{\rm cl}(\infty) \approx 0.6347$ in the even-$M$ case.
If $R \ll 1$, $\sigma_{\rm cl}(\infty) = 2/\pi \approx 0.6366$
with no parity dependence.
We show below that $\sigma_{\rm cl}$ is
the lower limit of the average conductivity.

Now, we present the numerical result
for the average conductivity $\langle\sigma\rangle$.
The system size is varied from $M \times N = 30 \times 10$ to $270 \times 90$
in the even-$M$ case and from $33 \times 11$ to $303 \times 101$
in the odd-$M$ case, where the aspect ratio is fixed at $R = 1/3$.
The number of impurities is also fixed as $N_{\rm imp} = M \times N$,
resulting in $\Gamma  = (a/v)^{2}V_{0}^{2}$.
The strength of disorder is tuned as $\Gamma = 0.2$--$8.0$
by adjusting $V_{0}$.
In performing the ensemble average,
the number of samples, $N_{\rm sam}$, is set to $5000$ for each data point.
The $L$ dependence of $\langle\sigma\rangle$ is shown in Fig.~3(a),
where open (filled) symbols correspond to the odd-$M$ (even-$M$) case,
and the dashed (solid) line represents $\sigma_{\rm cl}(L)$
in the odd-$M$ (even-$M$) case.
The strength of disorder is $\Gamma = 0.2$, $0.4$, $0.5$, $0.6$,
$0.8$, $1.0$, $1.2$, $1.5$, and $2.0$ from bottom to top.
If the numerical uncertainty $\Delta\sigma$ is defined as
$\Delta\sigma \equiv ({\rm var}\{\sigma\}/N_{\rm sam})^{1/2}$
following Ref.~\citen{markos},
the relative uncertainty $\Delta\sigma/\langle\sigma\rangle$
is smaller than $0.003$ at each data point.
No clear even-odd difference appears in Fig.~3(a).
We see that $\langle\sigma\rangle$ increases with increasing $\Gamma$,
indicating that $\langle\sigma\rangle$ is minimized in the clean limit,
i.e., $\langle\sigma(L)\rangle \ge \sigma_{\rm cl}(L)$.
Although $\langle\sigma\rangle$ slightly decreases with increasing $L$
at a small $\langle\sigma\rangle$,
this should not be regarded as a sign of localization.
Note that $\langle\sigma(L)\rangle$ at $\Gamma = 0.2$ (the lowest data set) is
very close to $\sigma_{\rm cl}(L)$ represented by the solid and dashed lines.
This indicates that the decreasing behavior is induced by
a finite-size correction reflecting a weak $L$ dependence of $\sigma_{\rm cl}$.
To reduce it, we introduce the renormalized conductivity
$\langle\langle\sigma(L)\rangle\rangle$ defined by
\begin{align}
  \langle\langle\sigma(L)\rangle\rangle
  = \langle\sigma(L)\rangle - \delta\sigma_{\rm cl}(L) ,
\end{align}
where the finite-size correction
\begin{align}
   \delta\sigma_{\rm cl}(L)
   \equiv \sigma_{\rm cl}(L) - \sigma_{\rm cl}(\infty)
\end{align}
is a monotonically decreasing function of $L$.
We treat $\langle\langle\sigma(L)\rangle\rangle$ instead of
$\langle\sigma(L)\rangle$ in the following analysis.
\begin{figure}[tbp]
\begin{minipage}{.47\linewidth}
\includegraphics[width=\linewidth]{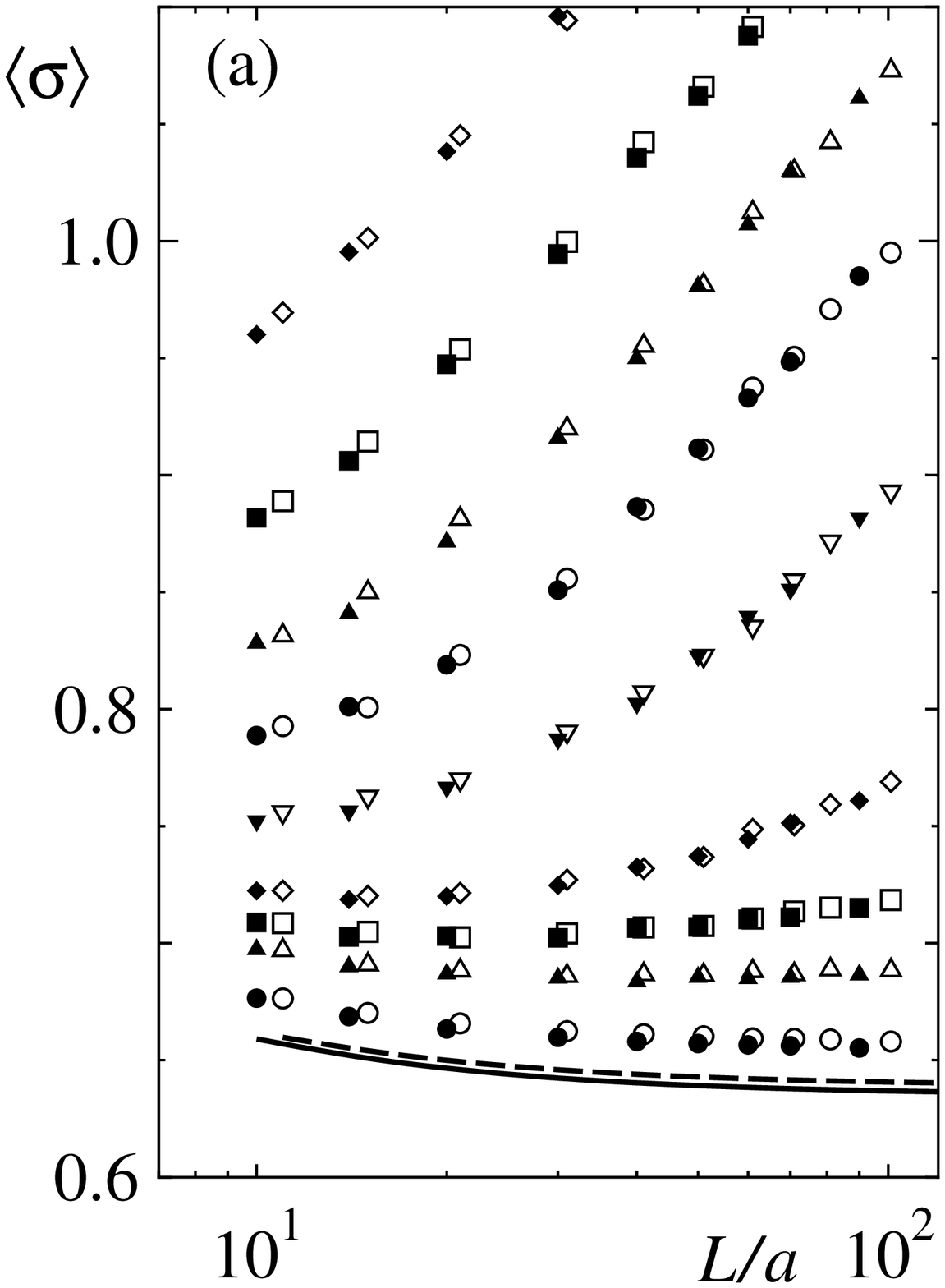}
\end{minipage}
\begin{minipage}{.47\linewidth}
\includegraphics[width=\linewidth]{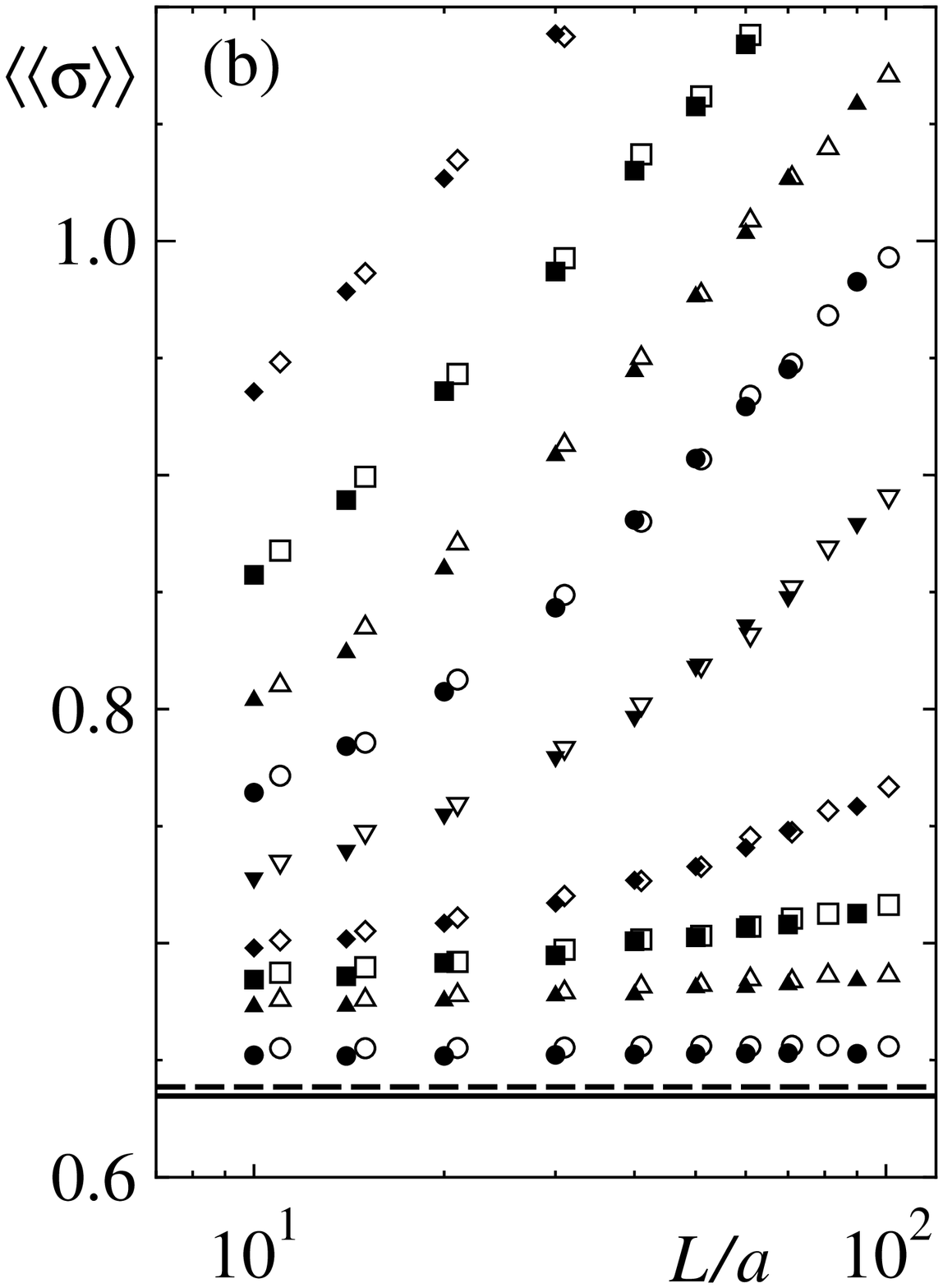}
\end{minipage}
\caption{(a) Average conductivity and (b) renormalized conductivity
as functions of $L$ for $\Gamma = 0.2$,
$0.4$, $0.5$, $0.6$, $0.8$, $1.0$, $1.2$, $1.5$, and $2.0$ from bottom to top.
Open (filled) symbols correspond to the odd-$M$ (even-$M$) case.
Dashed (solid) lines represent $\sigma_{\rm cl}$
in the odd-$M$ (even-$M$) case.
}
\end{figure}

The $L$ dependence of $\langle\langle\sigma\rangle\rangle$
is shown in Fig.~3(b).
We see that $\langle\langle\sigma\rangle\rangle$ clearly increases
with increasing $L$.
An exception is the case of $\Gamma = 0.2$ (the lowest data set),
where $\langle\langle\sigma\rangle\rangle$ seems almost independent of $L$.
However, this does not deny a plausible hypothesis that
$\langle\langle\sigma\rangle\rangle$ increases very slowly with increasing $L$
even when $\Gamma$ is small.
We demonstrate in Fig.~4 that the data sets for $\Gamma = 0.3$--$8.0$
collapse onto one scaling curve by shifting the data horizontally.
Here, only the data in the odd-$M$ case are plotted in Fig.~4.
We see that the scaling curve is approximated as
\begin{align}
     \label{eq:log-L}
  \langle\langle\sigma\rangle\rangle
  = {\rm const.} + \frac{1}{8}{\rm ln}(L/a^*) 
\end{align}
at a large $\langle\langle\sigma\rangle\rangle$,~\cite{comment2}
indicating the presence of
the weak antilocalization correction.~\cite{hikami}
We also see that the renormalized conductivity decreases toward
a limiting value, $\sigma^*$, with decreasing $L/a^*$.
Although $\sigma^*$ cannot be precisely determined from our data,
it is bounded from below by the clean-limit value $\sigma_{\rm cl}(\infty)$.
We adopt the most plausible conjecture that $\sigma^*=\sigma_{\rm cl}(\infty)$
since there is no reason to believe that $\sigma^*$ is larger than
$\sigma_{\rm cl}(\infty)$.
\begin{figure}[bpt]
\begin{center}
\includegraphics[height=5.8cm]{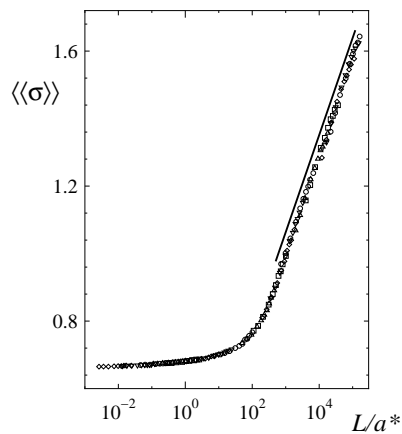}
\end{center}
\caption{One-parameter scaling plot of the renormalized conductivity,
where the data sets for $\Gamma = 0.3$, $0.325$, $0.35$, $0.375$, $0.4$,
$0.45$, $0.5$, $0.6$, $0.8$, $1.0$, $1.2$, $1.5$, $2.0$, $3.0$, $4.0$,
$6.0$, $7.0$, and $8.0$ are used.
The solid line represents Eq.~(\ref{eq:log-L}).
}
\end{figure}
The scaling function $\beta$, defined by
\begin{align}
  \beta\left(\langle\langle\sigma\rangle\rangle\right)
  = \frac{d{\rm ln}\langle\langle\sigma\rangle\rangle}{d{\rm ln}L} ,
\end{align}
is shown in Fig.~5.
Equation~(\ref{eq:log-L}) indicates that $\beta$
at a large $\langle\langle\sigma\rangle\rangle$
is expressed as $\beta = (1/8)\langle\langle\sigma\rangle\rangle^{-1} >0$.
In the opposite regime of $\langle\langle\sigma\rangle\rangle$ being
close to $\sigma_{\rm cl}(\infty)$, we expect that $\beta$ decreases
with decreasing $\langle\langle\sigma\rangle\rangle$
and finally vanishes at $\sigma_{\rm cl}(\infty)$.
Thus, $\beta$ is always positive except at $\sigma_{\rm cl}(\infty)$.
This is in marked contrast to the case of ordinary 2D systems
with symplectic symmetry,~\cite{markos,hikami,asada}
where $\beta$ becomes negative when the conductivity falls below
a critical value.
\begin{figure}[bpt]
\begin{center}
\includegraphics[height=2.7cm]{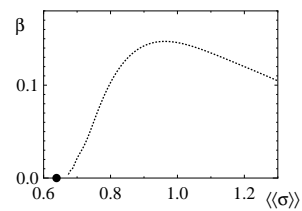}
\end{center}
\caption{Scaling function $\beta$,
where the filled circle corresponds to the clean limit.
}
\end{figure}

Here, let us briefly consider the case of $R$ being greater than $1/3$.
In this case, since the clean-limit conductivity $\sigma_{\rm cl}(\infty)$
depends on the even-odd parity as well as on $R$, we expect that
the scaling function at a small $\langle\langle\sigma\rangle\rangle$
will split into two curves corresponding to the even and odd cases.
We also expect that the two curves will merge
at a large $\langle\langle\sigma\rangle\rangle$
since the weak antilocalization correction should be insensitive to the parity.

The above argument indicates that the conductivity of the system monotonically
increases with increasing system size, except at an unstable fixed point
corresponding to the clean limit.
It is conceivable that such an unstable fixed point
also exists in 2D massless Dirac electron systems with one Dirac cone.

\section*{Acknowledgment}

The author thanks K.-I. Imura, Y. Yoshimura, T. Ohtsuki, K. Kobayashi,
and H. Obuse for valuable discussions.
This work is supported by a Grant-in-Aid for Scientific Research (C)
(No. 24540375).


\begin{thebibliography}{99}
\bibitem{comment2} The prefactor $1/8$ differs from
the previously reported value of $1/\pi$.~\cite{mong}
This is probably due to the fact that we fix the Fermi level
at the Dirac point and consider only a low-conductivity regime.
A difference similar to this is also observed
in the case of one Dirac cone.~\cite{bardarson}
\end{thebibliography}
\end{document}